\documentclass[%
 reprint,
 amsmath,amssymb,
 aps,
]{revtex4-2}
\usepackage{amsthm}
\usepackage{arydshln}
\usepackage{extarrows}
\usepackage{subfigure}
\usepackage{epsfig}
\usepackage{txfonts}
\usepackage{amsfonts}
\usepackage{esint}
\usepackage{mathrsfs}
\usepackage{amsmath}
\usepackage{color}
\usepackage{hyperref}
\usepackage{marvosym}
\usepackage[marginal]{footmisc}
\usepackage{appendix}
\usepackage{graphicx}
\usepackage{dcolumn}
\usepackage{bm}

\begin{document}
\title[]{Robust Phase metrology with hybrid quantum interferometers against particle losses}
\author{X. N. Feng}
\affiliation{Information Quantum Technology Laboratory, International Cooperation Research Center of China Communication and Sensor Networks for Modern Transportation, School of Information Science and Technology, Southwest Jiaotong University, Chengdu 610031, China}
\author{D. He}
\affiliation{Information Quantum Technology Laboratory, International Cooperation Research Center of China Communication and Sensor Networks for Modern Transportation, School of Information Science and Technology, Southwest Jiaotong University, Chengdu 610031, China}
\author{L. F. Wei\footnote{E-mail: lfwei@swjtu.edu.cn}}
\affiliation{Information Quantum Technology Laboratory, International Cooperation Research Center of China Communication and Sensor Networks for Modern Transportation, School of Information Science and Technology, Southwest Jiaotong University, Chengdu 610031, China}
\begin{abstract}
   Entanglement is an important quantum resource to achieve high sensitive quantum metrology. However, the rapid decoherence of quantum entangled states, due to the unavoidable environment noise, result in practically the unwanted sharp drop of the measurement sensitivity. To overcome such a difficulty, here we propose a spin-oscillator hybrid quantum interferometer to achieve the desirable precise estimation of the parameter encoded in the vibrations of the oscillator. Differing from the conventional two-mode quantum interferometers input by the two-mode NOON state or entangled coherent states (ECS), whose achievable sensitivities are strongly limited by the decoherence of the entangled vibrational states, we demonstrate that the present interferometer, input by a spin-dependent two-mode entangled state, possesses a manifest advantage, i.e., the measurement sensitivity of the estimated parameter is not influenced by the decoherence from the spin-oscillator entanglement. This is because that, by applying a spin-oscillator disentangled operation, the information of the estimated parameter encoded originally in the vibrational degrees can be effectively transferred into the spin degree and then can be sensitively estimated by the precise spin-state population measurements. As consequence, the proposed hybrid quantum interferometer possesses a manifest robustness against the particle losses of the vibrational modes. Interestingly, the achieved phase measurement sensitivity can still surpass the SQL obviously, even if relatively large number of particle loss occurs in one of the two modes. The potential application of the proposed spin-oscillator hybrid quantum interferometer is also discussed.
\end{abstract}
\maketitle
\section{Introduction}
In recent years, quantum metrology utilizing various quantum resources to beat the standard quantum limit (SQL) has become one of the hot topics in quantum information processing. Typically , quantum entanglement, as an important quantum resource, has been widely used to improve the measurement sensitivity of quantum interferometer. For example, if the Mach-Zender interferometer (MZI) is input by the NOON state, the achieved phase measurement sensitivity can reach the Heisenberg limit (HL)~\cite{Nagata,Yonatan,Boto} . In fact, quantum interferometers with NOON state inputs have been widely used to implement the super-resolution quantum lithography~\cite{Boto}, quantum microscopy~\cite{Yonatan} and the bio-sensing~\cite{Taylor, Anisimov} etc. However, a practical problem for these applications is that, the NOON state is inherently very fragile in various noise environments. Typically, with the loss of the photons the coherence of the interferometer decreases rapidly, leading to the achieved measurement sensitivity decays quickly~\cite{Zhang,Jaewoo,Huang,Oh,Dowling}.

To overcome such a practical problem, a series of proposals have been proposed~\cite{Jaewoo} to the MZIs typically including the squeezed state inputs~\cite{Caves,Pezz,Ma,Li,Zuo, Liu}), rather than the NOON state, to improve the robustness again the photon loss. However, the generation and manipulation of the strong squeezing states is still a great challenge for the current experimental technology. Therefore, the MZI input by either the ECS or the Schr\"{o}dinger-cat-states has become a hot research topic recently~\cite{Jaewoo}. Although these states are indeed more robust than the NOON state with respect to the noise of particle loss~\cite{Jaewoo,Huang}, the off-diagonal elements of the density matrix for these states, which provide the quantum enhancement of sensitivity beyond the SQL, still decays exponentially rapidly with the loss of particles~\cite{Zhang}. Therefore, it is still a challenge to achieve the high metrologic sensitivity far beyond the SQL with these states especially for large particles number. In addition, a detector with particle number resolved is usually required to achieve the high sensitivity detection beyond SQL. This is also a big challenge especially when the number of particles is large. Therefore, designing the novel quantum interferometer that is robust to the typical noise of particle loss as well as easy for readout is still an important issue.

Recently, the spin-oscillator hybrid quantum interferometer, a natural generalization of the MZI, have attracted much attention for quantum metrology~\cite{Pirkkalainen, Rugar, Kolkowitz, Ovart, Lee2}. Compared with the conventional two-mode interferometers, the spin-oscillator hybrid interferometers can well combine the advantages of the spin system and the oscillator one; the quantum harmonic oscillator is very sensitive to the changes of external environmental parameters, and thus they can be used as the ideal probes to achieve sensitive detection; the spin states are relatively easier to be manipulated, and thus the spin degrees of freedom can be used to achieve the high fidelity manipulation and readout of the oscillator states. In fact, these spin-oscillator hybrid quantum systems have been experimentally realized with various platforms, typically e.g., the trapped ions~\cite{Lo,Gilmore}, superconducting circuits~\cite{Kolkowitz,Pirkkalainen}, optomechanics~\cite{Aspelmeyer} and the diamond color centers~\cite{Lee2,Rugar,Kolkowitz} etc. In particular, in a trapped ion system, the coupling between the spin (i.e., the internal atomic levels) of the ion and its external vibration can be achieved by applying the designable laser pulses~\cite{Mizrahi}.
The parameter information encoded in the external state of the ion (such as the displacement amplitude caused by an external weak force) can be indirectly measured by probing the population of the ionic internal spin states~\cite{Gilmore, Ivanov}. Physically, the detection sensitivity of the spin state is limited by its deccoherence, due to the interaction with the harmonic oscillator~\cite{Campbell}. It was shown that, the spin decoherence of spin-oscillator interferometers can be effectively suppressed by squeezing the vacuum fluctuation of the oscillator~\cite{Lo,Feng} . Furthermore, Ref.~\cite{Gilmore} showed that, by applying a reverse action to the harmonic oscillator before the spin state detection, the inter- and external states of the trapped ions can be completely decoupled.
As a consequence, the fidelity of the spin-state detection can be enhanced significantly.

A natural question is, are these suppression methods of the spin decoherence also applicable to the spin-oscillator hybrid interferometer with a two-dimensional oscillator? More interestingly, whether the practical precision measurements are robust against the unavoidable environmental noise? In the following we will give the positive answers to these questions. Takes a spin-oscillator hybrid quantum interferometer with a two-dimensional quantum harmonic oscillator as an example, we investigate in this paper how to achieve the desirable precise measurement of the rotation parameter $\Omega$ with a spin-dependent entangled state. Here, the two-dimensional harmonic oscillator is equivalent to a conventional two-mode interferometer, which is used to encode the information of the rotation parameter ($\Omega$); the spin is used as an auxiliary degree of freedom to generate the desirable input state of the interferometer and the readout of the parameter information by performing the spin population detection. With certain spin-dependent entangled states as input, the spin-oscillator entanglement can be completely decoupled by applying the proper spin-oscillator inverse operations, yielding the information encoded in the two-dimensional oscillator states are transformed as the relative phase of the spin states. Therefore, by performing the high fidelity spin-state detection, the high sensitive measurement of the parameter can be achieved. More importantly, compared with the conventional two-mode interferometer input with NOON state and ECS~\cite{Gilmore}, we demonstrate that the proposed hybrid quantum interferometer is more robust to the noise induced by particles loss of the oscillator.

The paper is organized as follows. In Sec.~\ref{s2}, we introduce a general model of the hybrid spin-oscillator interferometer input with a general spin-dependent entangled state, to implement the parameter estimation. In Sec.~\ref{s3}, we investigate to improve the measurement sensitivity of the interferometer by inputting a special NOON-like spin-dependent entangled state. The specific measurement of parameter $\Omega$ through the spin projective measurement is also discussed.  In Sec.~\ref{s4}, the performance of the interferometer is investigated when the particle loss occur in the vibrating mode of oscillator is considered.  In Sec.~\ref{s5}, we summarize our results and discuss their experimental feasibilities.

\section{Spin-two-mode-oscillator hybrid quantum interferometer}\label{s2}
\begin{figure}[htbp]
  \centering
  \includegraphics[width=0.45\textwidth]{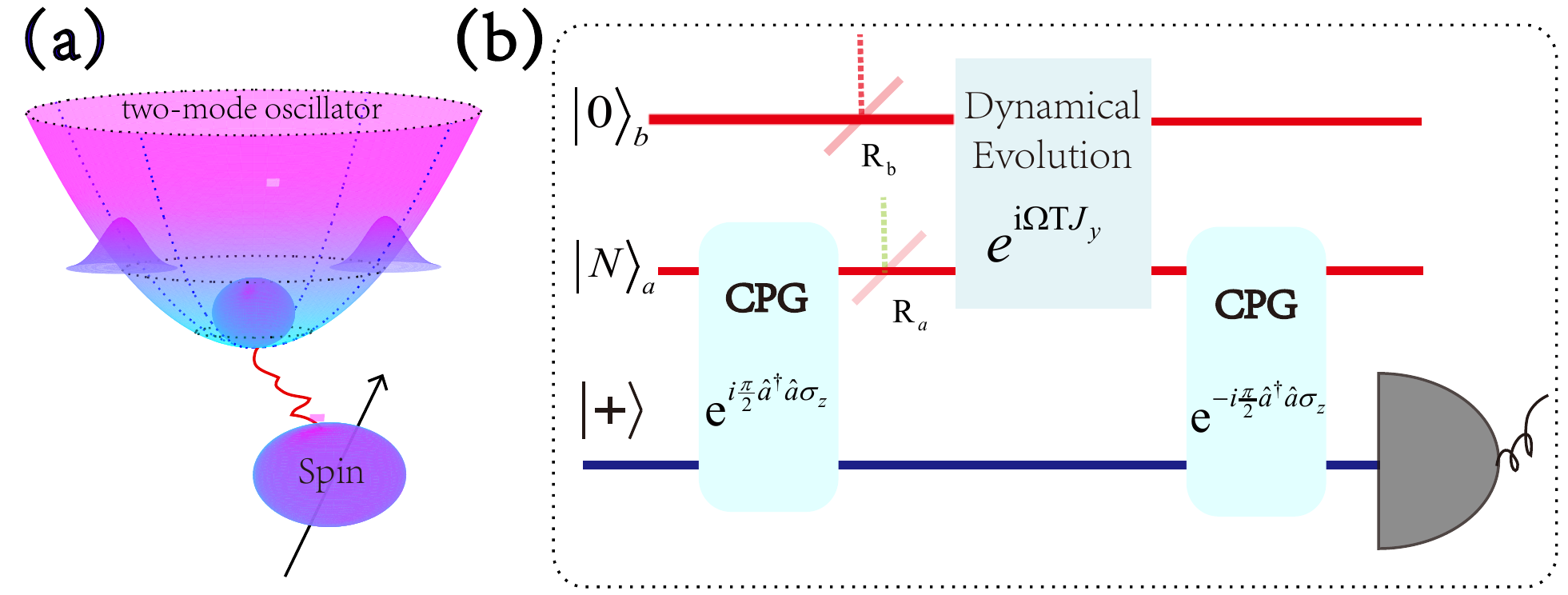}
  \caption{(Color online)A hybrid spin-oscillator quantum system (a) is used to implement the spin-oscillator quantum interferometer (b) with input state $|\Psi_i\rangle=|N\rangle_a|0\rangle_b\otimes|+\rangle$. A controlled-phase-gate (CPG) $e^{i\hat{a}^\dagger\hat{a}\sigma_z/2}$ and reverse CPG $e^{-i\hat{a}^\dagger\hat{a}\sigma_z/2}$ is applied before and after the phase accumulation $e^{i\Omega TJ_y}$ in the two-mode harmonic oscillator. $R_a$ and $R_b$ are the particle loss rates of the mode a and mode b, respectively. Finally, the information of the estimated parameter $\Omega$ is extracted by the projective measurement on the spin.} \label{syt}
\end{figure}

We consider a hybrid quantum interferometer system~\cite{Pirkkalainen, Rugar, Kolkowitz, Ovart, Lee2}, as shown schematically in Figure.~\ref{syt}, wherein a 1/2-spin is coupled to a two-dimensional harmonic oscillator. The parameter $\Omega$ to be estimated is encoded in the two-mode oscillator by the designed time-evolution with the time-independent Hamiltonian $\hat{H}=\Omega \hat{H}_0$, with $\hat{H}_0$ being the Hamiltonian of the oscillator. The spin is not only used to generate the desired spin-dependent entanglement of the hybrid quantum system, but also served as a detector to read out the information for the parameter estimations.
Without loss of generality, we assume that the hybrid quantum interferometer is prepared in the initial state (i.e, the input state of the interferometer):
\begin{equation}\label{eqs1}
  |\Psi_{i}\rangle=\frac{1}{\sqrt{2}}\Big[|\uparrow\rangle|\psi_1\rangle+|\downarrow\rangle|\psi_2\rangle\Big],
\end{equation}
where $|\psi_{1}\rangle$ and $|\psi_2\rangle$ denote the quantum states of the two-dimensional harmonic oscillator, $|\downarrow\rangle$ and $|\uparrow\rangle$ are the spin up and down states of the spin.
Obviously, if $|\psi_1\rangle=|\psi_2\rangle$, then the harmonic oscillator state and the spin state are separable and  unentangled. However, when $|\psi_1\rangle\neq\psi_2\rangle$, $|\Psi_i\rangle$ becomes a spin-dependent entangled state.

Supposed that the parameter $\Omega$  is encoded by the Hamiltonian $\hat{H}=\Omega\hat{H_0}$ of the system, which would evolve from the initial state $|\Psi_ {i}\rangle$ to the output state: $|\Psi(\Omega)\rangle=e^{i\Omega T\hat{H}_0}|\Psi_i\rangle$ with $T$ being the free evolution time of the system. As a consequence, the precise measurement of the parameter $\Omega$ can be realized by detecting the output state of the interferometer. According to the Cramer-Rao formula~\cite{Rao}, the measurement sensitivity theoretically satisfies the inequality:
\begin{equation}
  \Delta\Omega\geqslant \frac{1}{\sqrt{nF^Q(|\Psi_i\rangle)}}.
\end{equation}
Here, $n$ represents the number of measurements and $F^Q(|\Psi_i\rangle)$ is the QFI of the state $|\Psi_i\rangle$, which can be calculated as:
\begin{equation}\label{QFI}
  \begin{split}
  F^Q(|\Psi_{i}\rangle)&=4\Big(\langle\Psi_{i}|\hat{H}_0^2|\Psi_{i}\rangle-|\langle\Psi_{i}|\hat{H}_0|\Psi_{i}\rangle|^2\Big)\\
  &=F^Q_O(|\Psi_{i}\rangle)+F^Q_E(|\Psi_{i}\rangle),
  \end{split}
\end{equation}
with $F^Q_O(|\Psi_{i}\rangle)=2\Big[(\Delta\hat{H} _{0|\psi_1\rangle})^2+(\Delta\hat{H}_{0|\psi_2\rangle})^2\Big]$ and $F^Q_E(|\Psi_{i}\rangle)=\Big|\langle\psi_1|\hat{H}_0|\psi_1\rangle-\langle\psi_2|\hat{H}_0|\psi_2\rangle\Big|^2$, also $(\Delta\hat{H}_{0|\psi_j\rangle})^2=\langle\psi_j|\hat{H}_0^2|\psi_j\rangle-\langle\psi_j|\hat{H}_0|\psi_j\rangle^2$, $j=1,2$. Obviously, for $|\psi_1\rangle=|\psi_2\rangle$, we have $F^Q(|\Psi_i \rangle)=4 (\Delta\hat{H}_{0|\psi_1\rangle})^2$. This is result of the conventional two-mode quantum interferometer without spin entanglement. Interestingly, if $|\psi_1\rangle\neq|\psi_2\rangle$, the QFI of the state $|\Psi_{i}\rangle$ might increase, and thus the sensitivity of the parameter estimation can be improved. Specifically, suppose that~\cite{Campbell} $|\psi_1\rangle=|i\alpha\rangle_a|\beta\rangle_b$ and $|\psi_2\rangle=|-i\alpha\rangle_a|\beta\rangle_b$, with $|\pm i\alpha\rangle_a$ and $|\beta\rangle_b$ being respectively the coherent states of the mode a and moed-b, we have $(\Delta\hat{H}_{0|\psi_j\rangle})^2=|\alpha|^2+|\beta|^2$ $(j=1,2)$ and $\Big[\langle\psi_1|\hat{H}_0|\psi_1\rangle-\langle\psi_2|\hat{H}_0|\psi_2\rangle\Big]^2=4\Re(\alpha \beta)^2$, with $\hat{H}_0=J_y\equiv (\hat{a}^\dagger\hat{b}-\hat{a}\hat{b}^\dagger)/2i$. In particular, when $\alpha=|\alpha|=\beta\gg1$, the contribution of $F^Q_E(|\Psi_{i}\rangle)$ to QFI is much larger than that of $F^Q_O(|\Psi_i\rangle)$. In this case, the sensitivity of parameter estimation is mainly determined by the spin-oscillator entanglement.

As we know, with the NOON state or ECS as input the conventional two-mode interferometer without spin dependent can also achieve the precise measurements approaching the HL~\cite{Nagata,Jaewoo}. However, the single photon resolved detectors are usually required to achieve the desired sensitivity. For the present system, it is still a big challenge to realize the resolvable detection of the vibrational phonon numbers. To avoid such a difficulty, transferring the estimated parameters encoded in the harmonic oscillator states to the spin states, and then detecting it by the spin projection measurement is particularly desired. Basing on this idea, the precise measurement of parameter $\Omega$ encoded in the harmonic oscillator state can be accomplished by the following steps:

i) Let the spin-oscillator interferometer prepared in a spin-dependent entangled state undergos a dynamic evolution $\hat{U}(\Omega)$, i.e., $|\Psi_i\rangle\rightarrow|\psi(\Omega)\rangle=\hat{U}(\Omega)|\Psi_i\rangle$ for implementing the parameter encoding;

ii) Apply a global operation $\hat{F}=\exp(i\eta\hat{S}\otimes \hat{A})$ (with $\eta$ being the interaction strength between the spin and oscillator) on the state $|\psi(\Omega)\rangle$ for transferring the parameter into the spin state.
Here, $\hat{S}$ and $\hat{A}$ represent the operators acting on the spin state and the harmonic oscillator state, respectively. After such an operation, the state $|\psi(\Omega)\rangle$ of interferometer is evolved into:
\begin{equation}
|\psi_f\rangle=\hat{F}|\psi(\Omega)\rangle,
\end{equation}

iii) Perform the spin projection measurements after a $\pi/2$-pulse operation on the spin. The probability of the spin at the state $|\downarrow\rangle(|\uparrow\rangle)$ is obtained as:
\begin{equation}\label{population}
P_\downarrow(\Omega)=\langle\psi(\Omega)|\hat{F}^\dagger|\downarrow\rangle\langle\downarrow|\otimes I|\hat{F}|\psi(\Omega)\rangle,
\end{equation}
with $I$ being the identity operator of oscillator, $|\uparrow\rangle(|\downarrow\rangle)$ being the eigenstate of Pauli operator $\sigma_z$ with eigenvalue 1 (-1), and $P_\uparrow(\Omega)=1-P_\downarrow(\Omega)$.
Specifically, if $\hat{S}=\sigma_z$, then the equation~(\ref{population}) can be specifically expressed as:
\begin{equation}\label{p1}
P_\downarrow(\Omega)
=\frac{1}{2}\Big[1+\langle\psi_2|e^{-i\Omega TH_0}e^{2i\eta\hat{A}}e^{i\Omega TH_0}|\psi_1\rangle\Big].\\
\end{equation}
Obviously, if $\eta=0$ or $[\hat{A},\hat{H}_0]=0$, the spin measurement $P_\downarrow(\Omega)$ does not contain any information of the parameter $\Omega$, and thus such a measurement is invalid. While, for $[\hat{H}_0,\hat{A}]\neq0$, the result of the spin measurement must be a function of the parameter $\Omega$. Consequently, according to the error propagation formula, the sensitivity of the parameter estimation can be expressed as:
 \begin{equation}
   \Delta\Omega=\frac{\sqrt{P_\downarrow(\Omega)-P^2_\downarrow(\Omega)}}{d|P_\downarrow(\Omega)|/d\Omega}.
 \end{equation}
This indicates that, the sensitivity of the parameter estimation is closely related to the variation of $|P_\downarrow(\Omega)|$ with respect to the estimated parameter $\Omega$. For example, if $|\langle\psi_2^\prime|\psi_1^\prime\rangle|\neq1$ with $|\psi^\prime_{j}\rangle=e^{i(-1)^{ j+1}\eta\hat{A}}e^{i\Omega T\hat{H}_0}|\psi_j\rangle$ $(j=1,2)$, the value of  $|P_\downarrow(\Omega)|$ may decay with the increase of $\Omega T$, yielding the decrease of measurement sensitivity. Typically, as shown in Ref.~\cite{Campbell} that, if the spin-dependent cat state is utilized to measure the parameter $\Omega$, the spin measurement result
 \begin{equation}\label{p2}
  \tilde{P}_\downarrow(\Omega)
  =\frac{1}{2}\Big[1+e^{2i\alpha\beta\sin(\Omega T)}e^{-8\alpha^2\sin^2(\Omega T/2)}\Big],\\
  \end{equation}
decays exponentially with the increase of $\Omega T\leq\pi$, due to the spin state decoherence. Above, $\alpha$ and $\beta$ are the coherent displacement amplitudes of mode a and mode b, respectively. Obviously, in this case the high sensitivity of the parameter estimation can only be achieved for certain $\Omega T$ satisfying the condition $8\alpha^2\sin^2(\Omega T/2)\ll1$, such as $\Omega T\ll1$. Alternatively, in the following we prove that, by inputting a proper state to the interferometer and then performing the suitable quantum operations, the above spin decoherent effect induced by spin-oscillator entanglement can be effectively avoided. As a consequence, the desired precise measurements of the arbitrary $\Omega T$ parameter can be achieved with the proposed spin-oscillator hybrid quantum interferometer.

\section{Quantum metrology by using the hybrid interferometer with NOON-like state inputs}\label{s3}
From Eq.~(\ref{p1}) we can see that, the result of the spin state population measurement is related to the initial state, dynamic evolution, and the spin-oscillator interaction in the interferometer. Physically, the spin-oscillator entanglement might lead to the spin decoherence, which decreases the sensitivity of spin state population measurement. Recently, Gilmore et al.~\cite{Gilmore} has demonstrated an effective approach to avoid such a limit. For the weak force measurements with the one-dimensional vibration of a trapped ion crystal, they showed that the influence of the spin-oscillator entanglement induced decoherence on the population measurement can be effectively avoided by applying a inverse spin-dependent disentangled operation. In this section, we generalize such an idea to the proposed two-dimensional oscillator by similarly performing an inverse spin-oscillator operation to disentangle the spin and oscillator before the spin state population measurement.

For a typical parameter estimation model, wherein the dynamical evolution of the two-mode oscillator is described by the Hamiltonian $\hat{H}_0=J_y=(\hat{a}^\dagger\hat{b}-\hat{a}\hat{b}^\dagger)/2i$, suppose the hybrid interferometer is input with a spin-dependent entangled state
\begin{equation}\label{psi_i1}
\begin{split}
  |\Psi_{h}\rangle&=\hat{C}_a(\frac{\pi}{2})e^{i\frac{\pi}{2}J_y} |N,0\rangle|+\rangle\\
  &=\frac{(-i)^N}{\sqrt{2^{N+1} N!}}\Big[(\hat{a}^\dagger+i\hat{b}^\dagger)^N|0_a,0_b\rangle|\downarrow\rangle\\
  &+(-1)^N(\hat{a}^\dagger-i\hat{b}^\dagger)^N|0_a,0_b\rangle|\uparrow\rangle\Big],\\
  \end{split}
 \end{equation}
with $|N,0\rangle=|N\rangle_a|0\rangle_b$ and $|\pm\rangle=(|\uparrow\rangle\pm|\downarrow\rangle)/\sqrt{2}$ being the Fock state of the oscillator and the spin superposition state, respectively. Above, $\hat{C}_a(\pi/2)=e^{i\pi\hat{a}^\dagger\hat{a}\sigma_z/2}$ is a controlled-phase-gate (CPG)~\cite{Nemoto,Heuck,Sun}, acting on mode a of the oscillator, to generate the desired entangled state.
The estimated parameter $\Omega$ is encoded into the hybrid quantum interferometer under a free evolution:
\begin{equation}
  |\Psi(\Omega)\rangle=e^{i\Omega T J_y}|\Psi_h\rangle.
\end{equation}
The achievable sensitivity of its estimation is determined by the QFI of the output state $|\Psi(\Omega)\rangle$ of the interferometer. With Eq.~(\ref{QFI}), we have:
$\Delta J_{y|\varphi_k\rangle}=0$ and $\Big|\langle\varphi_1|J_y|\varphi_1\rangle-\langle\varphi_2|J_y|\varphi_2\rangle\Big|^2=N ^2$ with $|\varphi_k\rangle=e^{(-1)^{k+1}i\pi\hat{a}^\dagger\hat{a}/2}e^{i\pi/2 J_y}|N,0\rangle$ ($k=1,2$), and thus $F^{Q}(|\Psi(\Omega)\rangle)=N^2$. According to the Cramer-Rao inequality ~\cite{Rao}, the sensitivity of the estimated parameter can approach the HL. To implement the desired parameter estimation by using the spin population measurement, we apply an inverse CPG $\hat{C}^\dagger_a(\pi)$ to the output state, and obtain
\begin{equation}\label{psi_f}
\begin{split}
|\Psi_f\rangle&=\hat{C}^\dagger_a(\pi/2)|\Psi(\Omega)\rangle\\
&=e^{-i\pi\hat{a}^\dagger\hat{a}\sigma_z/2}e^{i\theta J_y}e^{i\pi\hat{a}^\dagger\hat{a}\sigma_z/2}e^{i\pi J_y/2}|N,0\rangle|+\rangle\\
&=\frac{1}{\sqrt{2}}(e^{i N\theta/2 }|\downarrow\rangle+e^{-i N\theta/2}|\uparrow\rangle)e^{i\pi J_y/2}|N,0\rangle,
\end{split}
\end{equation}
by which the parameter information is transferred into the relative phase of the spin state. Specifically, the population of the spin state $|\downarrow(\uparrow)\rangle$ reads:
\begin{equation}\label{p_1}
P_{\downarrow(\uparrow)}(\theta)=\frac{1}{2}\Big[1\pm\cos(N\theta)\Big],
\end{equation}
which implies that the sensitivity to estimate the parameter $\Omega$ can be expressed as
\begin{equation}\label{domega}
  \Delta\Omega=\frac{1}{NT}.
\end{equation}
Eqs.~(\ref{p_1}) and ~(\ref{domega}) show clearly that, the spin population oscillates rapidly with the phase $\theta=\Omega T$ and the higher sensitivity of the parameter estimation can be achieved by using the Fock state $|N\rangle$ with the larger phonon number of mode a.

Noted that, although the conventional two-mode interferometer with the NOON state input could also be utilized to implement the precise measurement approaching the HL, the required phonon number-resolvable detection is a challenge to be realized in practice. More importantly, the sensitivity reduces significantly with the loss of photon number, as the input NOON state is very fragile in the realistic environment. In the next section we show that, the interferometer proposed above is very robust against the phonon loss under certain condition, alternatively.

\section{Robustness for the "Particle" losses}\label{s4}
Physically, an entangled quantum system is usually fragile and its coherence losses easily under the disturbance of the environmental noise~\cite{Jaewoo}. Due to this decoherence, the sensitivity of the parameter estimation achieved by the interferometer with entangled state input would decreases. For example, the photon loss affects significantly the sensitivity of the typical two-mode interferometers~\cite{Jaewoo,Dowling}. For the spin-oscillator hybrid quantum interferometer proposed here, the vibrational phonon loss of the oscillator is also one of the main noises and lead to the decoherence of the hybrid quantum system. This will significantly reduce the achievable sensitivity of the parameter estimation.

Physically, the particle (e.g., phonon) loss of the spin-oscillator hybrid interferometer can be described by a beam-splitter~\cite{Oh}, wherein the noisy environment can be represented as a thermal bath in vacuum state $|0_E\rangle$. As a consequence, the input state~(\ref{psi_i1}) of the hybrid quantum interferometer should be replaced as:
\begin{equation}\label{loss_1}
\begin{split}
|\Psi\rangle_l &=\exp[\eta_a(\hat{a}^\dagger\hat{c}-\hat{a}\hat{c}^\dagger)]\exp[\eta_b(\hat{b}^\dagger\hat{e}-\hat{b}\hat{e}^\dagger)]|\Psi_h\rangle|0_E\rangle,
\end{split}
\end{equation}
where $\hat{c}^\dagger(\hat{c})$ and $\hat{e}^\dagger(\hat{e})$ are the creation (annihilation) operators of environment interacting with the mode a and mode b of the oscillator, respectively. $R_a=\sin^2(\eta_a)$ and $R_b=\sin^2(\eta_b)$ are the corresponding loss rates. Considering these particle losses, the equation~(\ref{loss_1}) can be rewritten as (see Appendix~\ref{A} )
\begin{equation}\label{psi_l}
   |\Psi\rangle_l
   =\frac{\hat{C}_a(\pi/2)e^{i\pi J_y/2}}{\sqrt{N!}}(u\hat{a}^\dagger+v\hat{b}^\dagger+p\hat{e}+q\hat{c}\sigma_z)^N|0_a0_b,0_E\rangle|+\rangle,
\end{equation}
where $u=(\cos(\eta_a)+\cos(\eta_b))/2$, $v=(\cos(\eta_a)-\cos(\eta_b))/2$, $p=i \sin(\eta_a)/\sqrt{2}$, and $q=\sin(\eta_b)/\sqrt{2}$. After tracing the wave function of the environment and the harmonic oscillator, the reduced density matrix of the spin state is obtained as (see Appendix~\ref{B} for detail)
\begin{equation}\label{rho_L}
  \rho_L=\frac{1}{2}\Big(|\downarrow\rangle\langle\downarrow|+|\uparrow\rangle\langle\uparrow|\Big)+R|\downarrow\rangle\langle\uparrow|+R^*|\uparrow\rangle\langle\downarrow|,
\end{equation}
It is seen that the information of the parameter $\theta=\Omega T$ is related to the off-diagonal element
\begin{equation}\label{nde}
R=(u^2e^{i\theta}+v^2e^{-i\theta}+p^2-|q|^2)^N,
\end{equation}
of the above reduced density matrix.
Similar to the Ramsey interference, after a $\pi/2$-pulse operation the measured result of the spin state population is:
\begin{equation}\label{pl}
  P^L_{\downarrow}(\theta)=\frac{1}{2}\Big[1+|R|\cos(\phi(\theta))\Big],\,\phi(\theta)=\arg(R),
\end{equation}
Obviously, with the error propagation formula, the estimated sensitivity of the $\theta$-parameter can be expressed as
\begin{equation}\label{dtheta_l}
  \begin{split}
    \Delta\theta= \frac{\sqrt{P^L_{\downarrow}-(P^L_{\downarrow})^2}}{|dP^L_{\downarrow}/d\theta|}
    \approx\frac{\sqrt{1-|R|^2\cos^2(\phi(\theta))}}{|R||\sin(\phi(\theta))d\phi(\theta)/d\theta|}.
  \end{split}
\end{equation}
\begin{figure}
  \centering
   \includegraphics[width=0.45\textwidth]{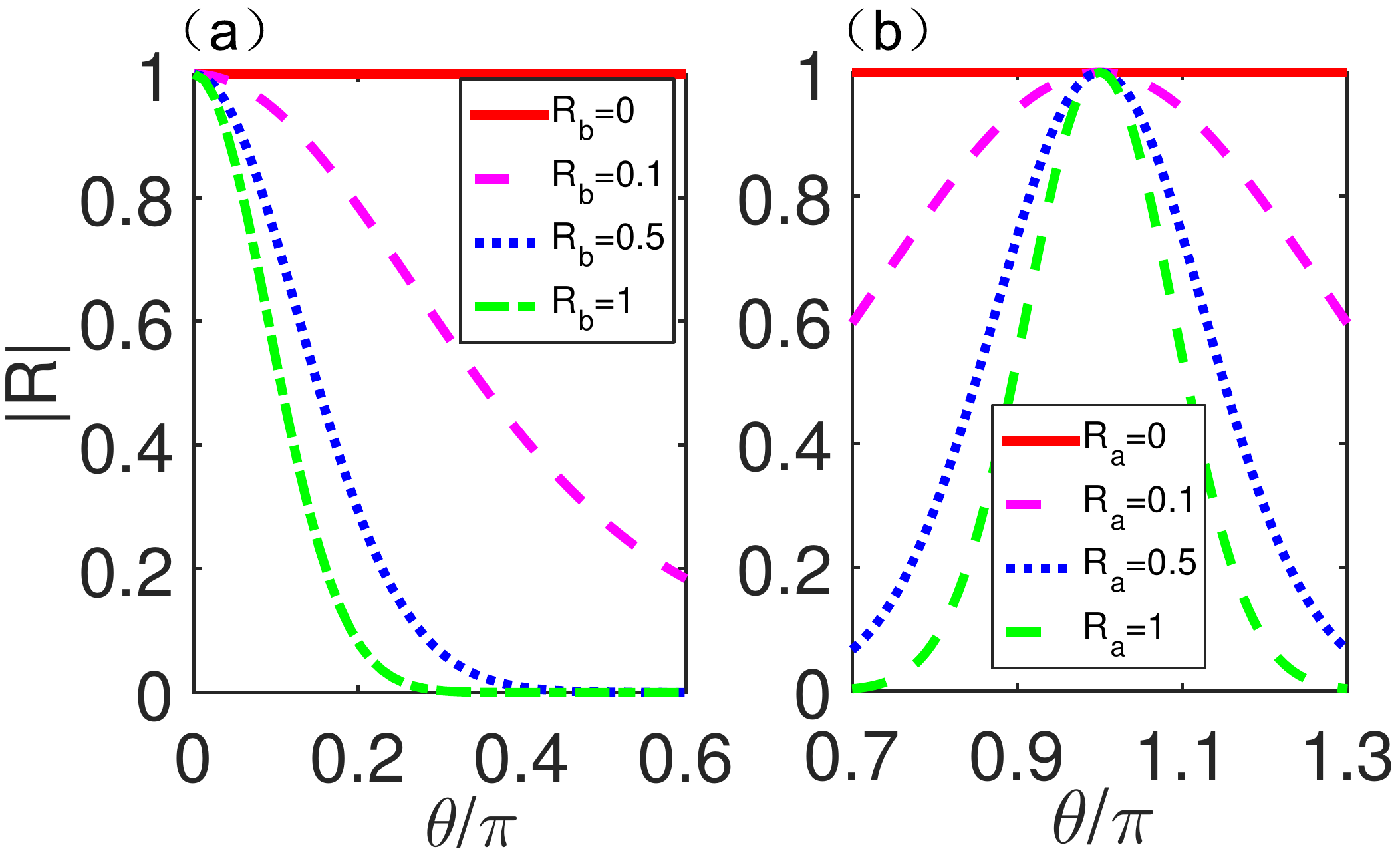}
\caption{(Color online) The non-diagonal element of the reduced matrix of spin $|R|$ versus the phase $\theta$ for the particle loss occurring only in one mode (mode b (the left subfigure) or mode a(the right subfigure)) with different loss rates. }\label{g_c}
\end{figure}
It is seen from Eq.~(\ref{dtheta_l}) that the sensitivity $\Delta\theta$ is mainly determined by the value of the off-diagonal element $R$ (i.e., the value of $|R|$ and $\phi(\theta)=\arg(R)$). Obviously, when $|R|=1$ the spin state $\rho_L$ of Eq.~(\ref{rho_L}) is a pure state, and the sensitivity of Eq.~$(\ref{dtheta_l})$ depends only on the derivation $d\phi(\theta)/d\theta$ and becomes $\Delta\theta\approx1/|d\phi(\theta)/d\theta|$. While $|R|<1$, the state $\rho_L$ becomes a mixed state, and the estimation sensitivity $\Delta\theta$ decreases.
\begin{figure}[htbp]
  \centering
  \includegraphics[width=0.45\textwidth]{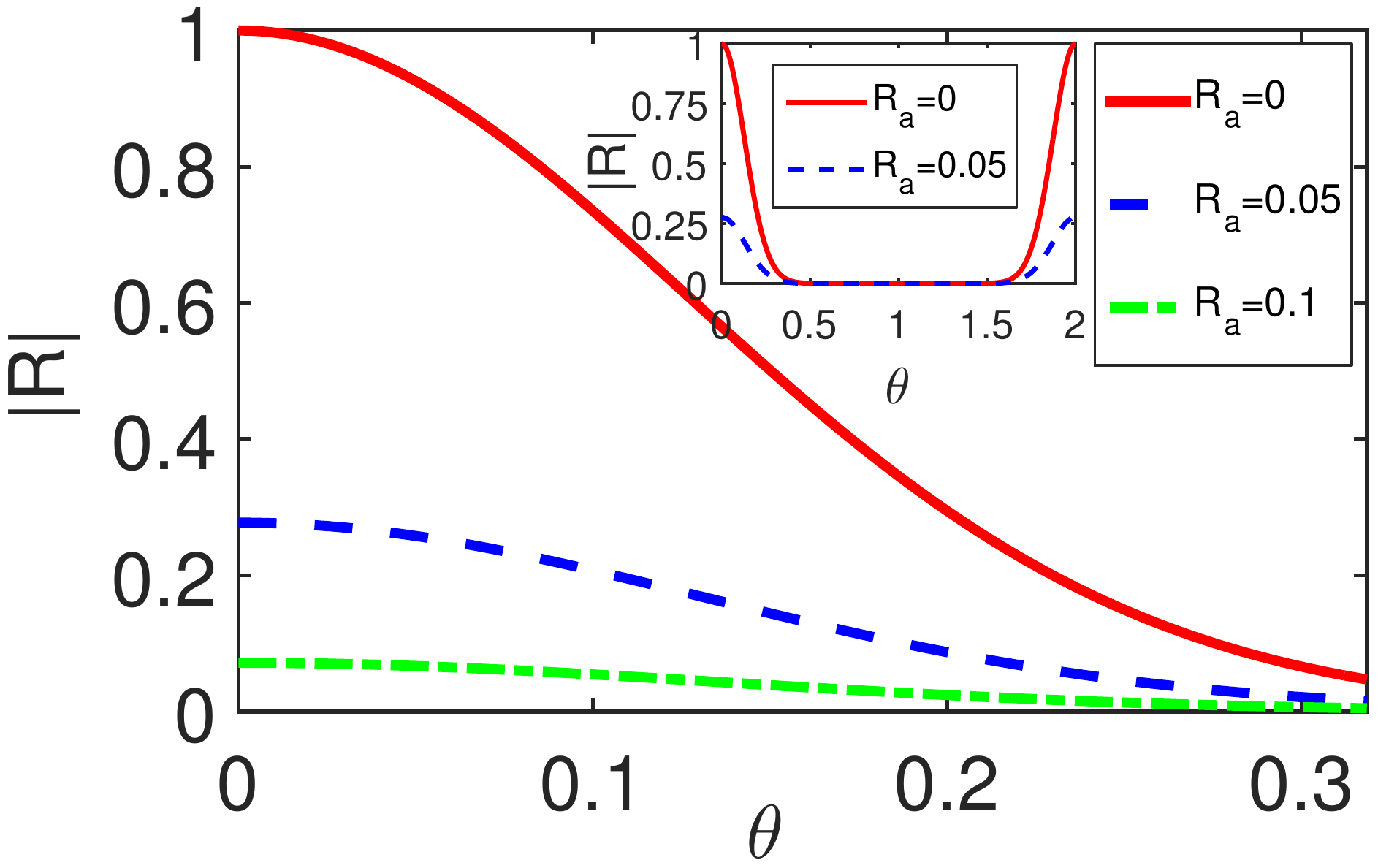}
  \caption{(Color online) The non-diagonal element $|R|$ of the reduced density matrix of spin versus the phase $\theta$ under the different particle loss rates of mode-a. Here, the loss rate for the mode b is fixed as: $R_b=0.5$, for $N=25$.}\label{g_ab}
\end{figure}
Here the value of $|R|$ are related to the loss rate $R_a$, $R_b$ and the value of $\theta$ as shown in Eq.~(\ref{nde}). Fig.~\ref{g_c} shows how the value of the off-diagonal element for the spin state varies with the phase $\theta$ for single mode loss (either mode a or mode b). It is shown that, the value of $|R|\approx1$ is achieved around $\theta=0$ (or $\theta=\pi$) as long as the particles loss only occurs in mode b with $R_a=0$ (or mode a for $R_b=0$). Though with increase of loss rate the value of $|R|$ decays more rapidly with $\theta$, $|R|\approx1$ can still be achieved. This implies that the high sensitivity can still be achieved in this case as shown below. While the particle loss occur in both modes the value of $|R|(\leq 1)$ also drop quickly and can not achieve one again for any $\theta$ as shown in Fig.~\ref{g_ab}. Given $\theta$ is precisely determined, the value of $\Omega$ can be estimated finally with the sensitivity being $\Delta\Omega=\Delta\theta/T$ for a fixed evolution time $T$.

Noted that, the influence of different loss rates for two modes on the estimated sensitivity of the phase achieved by the conventional two-mode interferometers have been widely investigated~\cite{Zhang,Jaewoo,Dobrzanski,Lee,Kacprowicz,Ulanov}. For example, the Refs~\cite{Zhang, Dobrzanski} have proven that in a two-mode interferometer with either NOON state or ECS input, the sensitivity of the parameter estimation is mainly determined by the larger loss rate of the two mode. Therefore, as long as one mode loses relatively large particle number, the achieved sensitivity of the estimation would rapidly decrease regardless of whether the other mode occurs the particle loss.
Alternatively, with the hybrid quantum interferometer proposed here, the optimal sensitivity achieved by using this hybrid system is mainly determined by the smaller loss rate since the value of $|R|$ mainly depends on the smaller loss rate shown in Fig.~\ref{g_ab}.
\begin{figure}[htbp]
  \centering
  \includegraphics[width=0.45\textwidth]{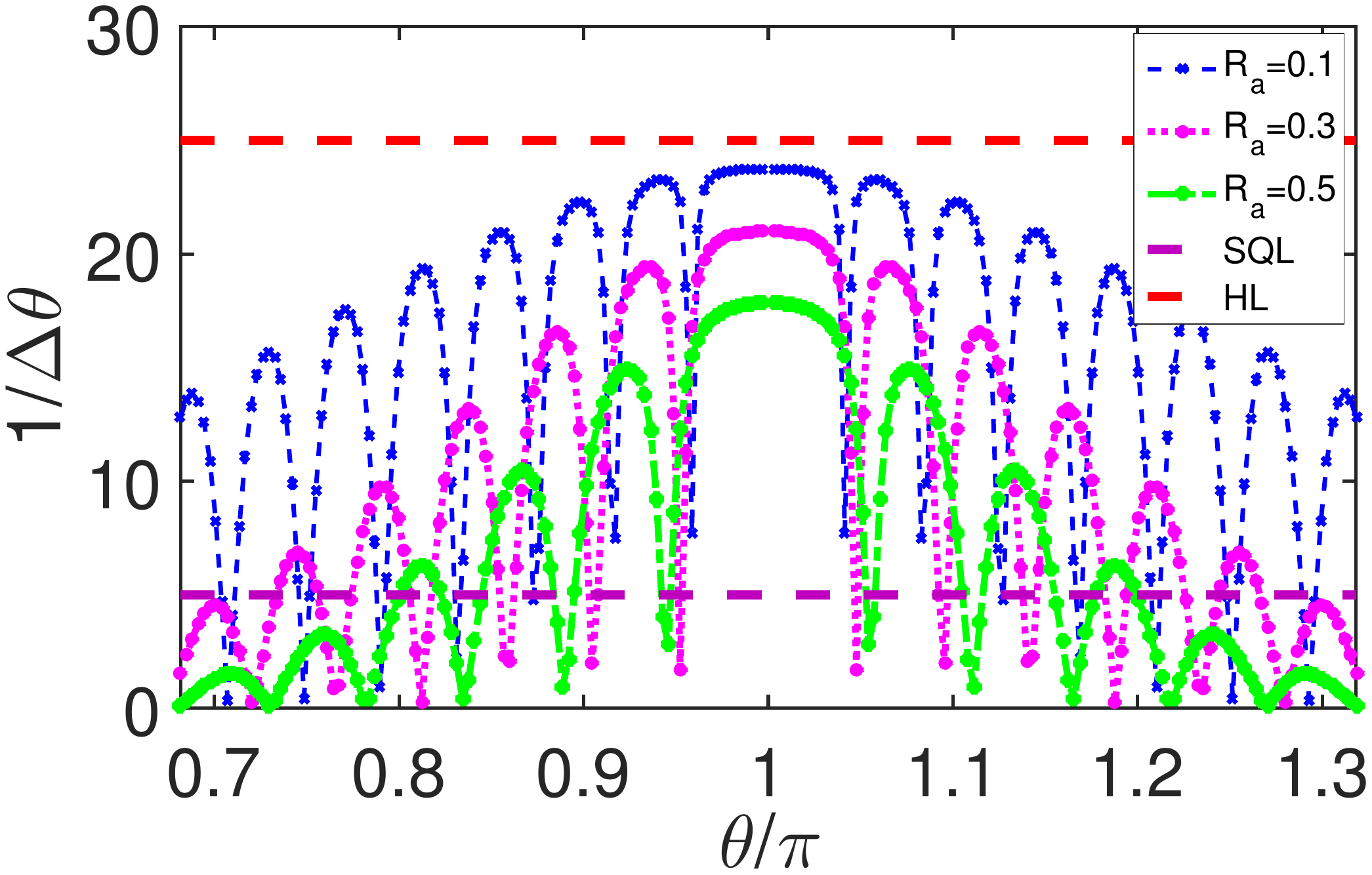}
  \caption{(Color online) The sensitivity of the phase estimation, characterized by $1/\Delta\theta$, changes with the phase $\theta$ for various loss rates of the mode-a. Here, $R_b=0$ for $N=25$.}\label{dtheta}
\end{figure}
Fig.~\ref{dtheta} shows how the variation of the phase measurement sensitivity, characterized by $1/\Delta\theta$ in y-axis, changes with the phase $\theta$ for different loss rates. Obviously, the larger the value of $1/\Delta\theta$ corresponds to the higher achievable sensitivity of the phase estimation. It can be seen that, if particle loss only occur in mode b, the optimal sensitivity of the phase estimation can be achieved at $\theta\approx0$. Typically, even if the loss rate of the mode b is relatively large, approaching to such as $50\%$, the optimal sensitivity still obviously surpasses the SQL.
\begin{figure}[htbp]
  \centering
  \includegraphics[width=0.45\textwidth]{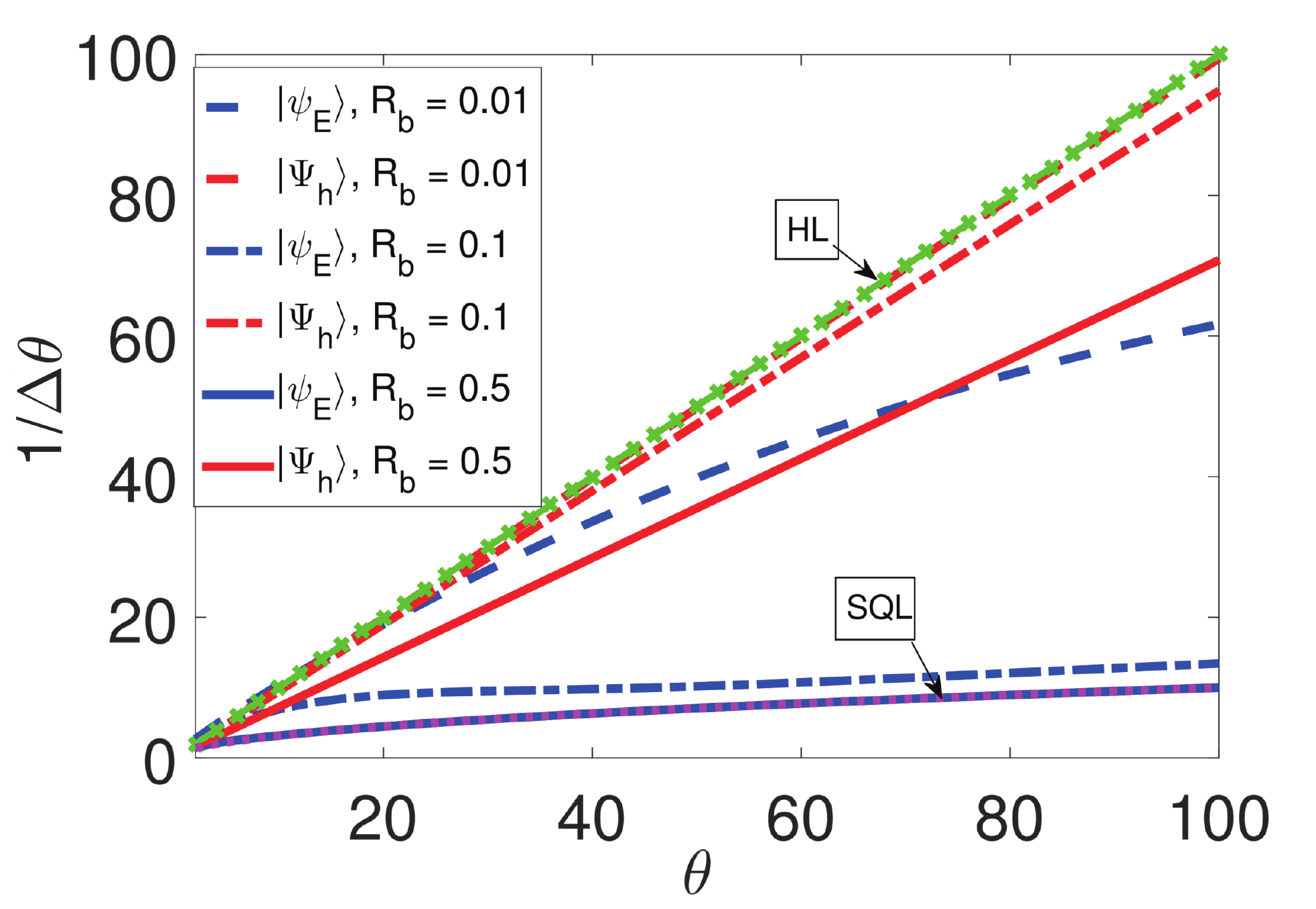}
  \caption{(Color online) The sensitivity, characterized by $1/\Delta\theta$, achieved by
  the present hybrid interferometry (marked by its input state $|\Psi_h\rangle$) compared with that demonstrated by the convenient interferometer (typically in Ref.~\cite{Lee}) marked as $|\Psi_E\rangle$ in the figure. Here, we assume that the particle loss occurs only in mode b with the decay rate $R_b$. }\label{dhe}
\end{figure}

To show more clearly the loss robustness of the present hybrid interferometer input by $|\Psi_h\rangle$, let us compare its reachable sensitivity with that typically demonstrated with the conventional two-mode interferometer input by the ECS~\cite{Lee}. Fig.~\ref{dhe} shows that, if the particle loss occurs only in one mode with the same rate, the sensitivity achieved by the present hybrid interferometer is much higher than that achieved by the conventional two-mode interferometer. Furthermore, even for the relatively small loss rate, such as $R=0.1$, the sensitivity with ECS quickly decreases to the SQL, while the sensitivity with $|\Psi_h\rangle$ is still significantly high approaching to the HL. Even if the loss rate is relatively large such as $R_b=0.5$, the sensitivity with $|\Psi_h\rangle$ is still about seven times larger than the SQL, typically for $N=100$ and $\theta=0.01$.

\section{Conclusions and Discussions}\label{s5}
In summary, an entangled hybrid quantum interferometer is proposed to implement the quantum precision measurement of the parameters encoded in a two-dimensional oscillator.
We showed that, the parameter wanted to be estimated can be completely transferred into the spin state of the interferometer by using the spin-oscillator interaction. After a free evolution, another spin-oscillator disentangled operation is applied to transfer the information into the spin state for the spin-state population measurement. The significant advantage of this precision measurement scheme is, the desired spin state population measurements can avoid the influence from the spin-oscillator entanglement. Therefore, the achievable sensitivity of the parameter estimation can be significantly high, even if one of the vibrational mode of the oscillator exists the dissipated due to the particle loss. By comparing with the conventional spin-free two-mode bosonic interferometers, wherein the loss of the particles in any mode would lead to the rapid decrease of the sensitivity, we showed that the spin-oscillator hybrid quantum interferometer input with certain spin-dependent entangled state proposed here is more robust to the particle losses.

Given a series of experimental platforms (typically such as the well-known cooled trapping ions, the vibrational spin ensembles and also the electrons trapped on liquid Helium, etc.,~\cite{Marcos,Zhu,Gangloff, Anders, Huang}) have demonstrated the interaction between the two-level atoms (spins) and the two-dimensional bosonic vibrations, and also the successful preparations of the desired vibrational Fock states~\cite{McCormick,Wolf}, we believe that the quantum metrology with the proposed hybrid interferometers is feasible. Also, the robustness of the proposal interferometer against the particle losses in the vibrations of the two modes can also be discussed similarly.

\section{Acknowledgments}
This work is partially supported by the National Natural Science Foundation of China (NSFC) under Grant No. 11974290, and the National Key Research and Development Program of China (NKRDC) under Grant No. 2021YFA0718803.

\begin{appendix}
\section{Derivation of the Eqs.~(\ref{psi_l})}\label{A}
In this appendix, we provide the relevant derivations by considering the effects of the population losses on measurement accuracy. Following in the conventional two-mode quantum interferometer, the particle loss can be described by the beam splitters of the mode a and mode b. Therefore, with the particle number loss, the input state of the interferometer can be written as:
  \begin{equation}
    \begin{split}
    |\Psi\rangle_l &=\exp[\sqrt{2}\eta_a(\hat{a}^\dagger\hat{c}-\hat{a}\hat{c}^\dagger)]\exp[ \sqrt{2}\eta_b(\hat{b}^\dagger\hat{e}-\hat{b}\hat{e}^\dagger)]|\Psi_h\rangle\\
    &=\hat{C}_a(\pi/2)e^{i\pi/2 J_y}\exp\Big[i\eta_a\sigma_z\big((\hat{a}^\dagger+\hat{b }^\dagger)\hat{c}+h.c.\big)\\
    &+\eta_b\big((\hat{b}^\dagger-\hat{a}^\dagger)\hat{e}-h.c.\big) \Big]|N,0\rangle|0_E\rangle|+\rangle,
    \end{split}
  \end{equation}
where $R_{k}=\sin^2(\eta_k)$ $(k=a,b)$ represents the loss rate of the particles of the mode a and mode b, respectively.
$\hat{C}_a(\pi /2)=\exp(i\pi\hat{a}^\dagger\hat{a}\sigma_z/2)$ with $|0_E\rangle=|0_c,0_e\rangle$ being the vacuum modes of the environment.
Using the Baker-Hausdorff formula~\cite{Menda},
  \begin{equation}
   e^{\hat{A}}\hat{B} e^{-\hat{A}}=\sum^\infty_{n=0}\frac{1}{n!}[\hat{A^ {(n)}},\hat{B}],
  \end{equation}
with $[\hat{A}^{(n)},\hat{B}]=[\hat{A},[\hat{A}^{(n-1)},\hat{B}] ]$ and $[\hat{A}^{(0)},\hat{B}]=\hat{B}$, and letting $\hat{A}=i\eta_a\sigma_z[(\hat{a}^\dagger+\hat{b}^\dagger)\hat{c}+h.c.]
  +\eta_b[(\hat{b}^\dagger-\hat{a}^\dagger)\hat{e}-h.c.]$, $\hat{B}=\hat{a}^\dagger$, we have
  \begin{equation}
    \begin{split}
     [\hat{A}^{(2n+1)},\hat{a}^\dagger]&=2^{n}(i\eta_a\sigma_z)^{2n+1}\hat{c}^ \dagger+(-1)^{n}2^{n}\eta_b^{2n+1}\hat{e}^\dagger\\
     [\hat{A}^{(2n)},\hat{a}^\dagger]&=2^{n-1}(i\eta_a\sigma_z)^{2n}(\hat{a}^\dagger+\hat{b}^\dagger)+(-1)^{n-1}2^{n-1}\eta_b^{2n}(\hat{b}^\dagger-\hat{a}^\dagger).
   \end{split}
  \end{equation}
  Consequently, we have
  \begin{equation}\label{A4}
  \begin{split}
  &e^{\hat{A}}\hat{a}^\dagger e^{-\hat{A}}=\sum_{n=0}^{\infty}\frac{1}{n!}[\hat{A}^{(n)},\hat{a}^\dagger]\\
  =&\sum_{k=0}\frac{(\sqrt{2}i\eta_a\sigma_z)^{2k}}{2(2k)!}(\hat{a}^\dagger+\hat{b}^\dagger)
  +\sum_{k=0}\frac{(-1)^{k+1}(\sqrt{2}\eta_b)^{2k}}{2(2k)!}(\hat{b}^\dagger-\hat{a}^\dagger)\\
  &+\sum_{k=0}\frac{(\sqrt{2}i\eta_a\sigma_z)^{2k+1}}{\sqrt{2}(2k+1)!}\hat{c}^ \dagger
  +\sum_{k=0}\frac{(-1)^k(\sqrt{2}\eta_b)^{2k+1}}{\sqrt{2}(2k+1)!}\hat{e }^\dagger\\
  =&\frac{1}{2}\Big[\cos(\sqrt{2}\eta_a)(\hat{a}^\dagger+\hat{b}^\dagger)-\cos(\sqrt{ 2}\eta_b)(\hat{b}^\dagger-\hat{a}^\dagger)\Big]\\
  &+\frac{1}{\sqrt{2}}i\sin(\sqrt{2}\eta_a)\sigma_z\hat{c}^\dagger+\frac{1}{\sqrt{2}}\sin (\sqrt{2}\eta_b)\hat{e}^\dagger\\
  =&\alpha\hat{a}^\dagger+\beta\hat{b}^\dagger+p\hat{e}^\dagger+q\hat{c}^\dagger\sigma_z,
  \end{split}
  \end{equation}
  where $\alpha=(\cos(\sqrt{2}\eta_a)+\cos(\sqrt{2}\eta_b))/2$, $\beta=(\cos(\sqrt{2}\eta_a)-\cos(\sqrt{2}\eta_b))/2$, $q=i\sin(\sqrt{2}\eta_a)/\sqrt{2}$ and $p=\sin (\sqrt{2}\eta_b)/\sqrt{2}$.
  By Noting that $e^{\hat{A}}|N,0\rangle |0_E\rangle=\Big[e^{\hat{A}}\hat{a}^\dagger e^{\hat{A}}\Big]^N/\sqrt{N!}|\Theta\rangle$ with $|\Theta \rangle=|0_a,0_b\rangle|0_E\rangle$ being the vacuum of the oscillator system and the environment and combining the above Eq.~(\ref{A4}), we get
  \begin{equation}
  \begin{split}
    \widetilde{|\Psi\rangle}_l=&\exp(\hat{A})|N,0\rangle|0_E\rangle|+ \rangle\\
    =&\frac{1}{\sqrt{N!}}(\alpha\hat{a}^\dagger+\beta\hat{b}^\dagger+p\hat{e}^\dagger+q\hat {c}^\dagger\sigma_z)^N|\Theta \rangle|+\rangle\\
    =&\frac{1}{\sqrt{N!}}\sum_{k=0}^NC_N^k\hat{S}^{ N-k} (p\hat{e}^\dagger+q\hat{c}^\dagger\sigma_z)^k|\Theta \rangle|+\rangle\\
    =&\frac{1}{\sqrt{2N!}}\sum_{k=0}^NC_N^k\hat{S}^{ N-k}(p\hat{e}^\dagger+q\hat{c}^\dagger)^{k}|\Theta \rangle|\downarrow\rangle\\
    &+\frac{1}{\sqrt{2N!}}\sum_{k=0}^NC_N^k\hat{S}^{ N-k}(p\hat{e}^\dagger-q\hat{c}^\dagger)^{k}|\Theta \rangle|\uparrow\rangle\\
    \equiv&\hat{D}|\Theta \rangle|\downarrow\rangle+\hat{U}|\Theta \rangle|\uparrow\rangle,
  \end{split}
  \end{equation}
  where $\hat{S}=\alpha\hat{a}^\dagger+\beta\hat{b}^\dagger$ and
  \begin{equation}\label{psi_la}
  \begin{split}
  |\Psi\rangle_l=&\hat{C}_a(\pi/2)e^{i\pi/2J_y}\widetilde{|\Psi\rangle}_l\\
  =&\frac{\hat{C}_a(\pi/2)e^{i\pi J_y/2}}{\sqrt{N!}}(u\hat{a}^\dagger+v\hat{b}^\dagger+p\hat{e}+q\hat{c}\sigma_z)^N|0_a0_b,0_E\rangle|+\rangle.
  \end{split}
  \end{equation}
  Such Eq.~(\ref{psi_l}) is obtained.
  \section{Derivation of the Eq.~(\ref{rho_L}) of the reduced density matrix of spin under particle loss}\label{B}
  The state $\widetilde{|\Psi\rangle}_l$ can be also represented as a density matrix:
    \begin{equation}
     \begin{split}
      \hat{\rho}
      =&\hat{D}|\Theta\rangle\langle\Theta|\hat{D}^\dagger\rho_{\downarrow\downarrow}
      +\hat{U}|\Theta\rangle\langle\Theta|\hat{U}^\dagger\rho_{\uparrow\uparrow}\\
      &+\hat{D}|\Theta\rangle\langle\Theta|\hat{U}^\dagger\rho_{\downarrow\uparrow}
      +\hat{U}|\Theta\rangle\langle\Theta|\hat{D}^\dagger\rho_{\uparrow\downarrow},
     \end{split}
   \end{equation}
   where $\rho_{\downarrow\downarrow}=|\downarrow\rangle\langle\downarrow|$,$\rho_{\uparrow\uparrow}=|\uparrow\rangle\langle\uparrow|$ and $\rho_{\downarrow\uparrow}=|\uparrow\rangle\langle\uparrow|=\rho^*_{\uparrow\downarrow}$
   Tracing out the environment variables, we get
    \begin{equation}
        \begin{split}
          \hat{\rho}_1=&\hat{\rho}_{00} \rho_{\downarrow\downarrow}
          +\hat{\rho}_{11} \rho_{\uparrow\uparrow}\\
          &+\hat{\rho}_{01} \rho_{\downarrow\uparrow}
          +\hat{\rho}_{10} \rho_{\uparrow\downarrow}
        \end{split}
      \end{equation}
     where $|0_s\rangle=|0_a,0_b\rangle$ is the ground state of the two-dimensional oscillator, and also
      \begin{equation}
        \begin{split}
        \hat{\rho}_{00} &= Tr_E\Big[\hat{D}|\Theta\rangle\langle\Theta|\hat{D}^\dagger\Big]\\
        &=\sum_{n=0,k=0}^N\frac{\hat{S}^{N-k}\text{Tr}_E\Big[(p\hat{e}^\dagger+q\hat{c}^\dagger)^{k}|\Theta\rangle\langle \Theta|(p^*\hat{e}+q^*\hat{c})^{n}\Big]\hat{S}^{\dagger(N-n)}}{2N!}\\
        &=\sum_{k=0}^N\frac{\hat{S}^{N-k}\text{Tr}_E\Big[(p\hat{e}^\dagger+q\hat{c}^\dagger)^{k}|\Theta\rangle\langle \Theta|(p^*\hat{e}+q^*\hat{c})^{k}\Big]\hat{S}^{\dagger(N-k)}}{2N!}\\
        &=\sum_{k=0}^N\sum_{m=0}^k\frac{(C_k^{m})^2|p|^{2m}|q|^{2k-2m}m!(k-m)!\hat{S}^{N-k}|0_s\rangle\langle 0_s|\hat{S}^{\dagger(N-k)}}{2N!}\\
        &=\frac{1}{2N!}\sum_{k=0}^Nk!(|p|^2+|q|^2)^k\hat{S}^{N-k}|0_s\rangle\langle 0_s|\hat{S}^{\dagger(N-k)},
        \end{split}
        \end{equation}
      and
      \begin{equation}
        \begin{split}
        \hat{\rho}_{01} &= Tr_E\Big[\hat{D}|\Theta\rangle\langle\Theta|\hat{U}^\dagger\Big]\\
        &=\sum_{k=0,n=0}^N\frac{\hat{S}^{N-k}\text{Tr}_E\Big[(p\hat{e}^\dagger+q\hat{c}^\dagger)^{k}|\Theta\rangle\langle \Theta|(p^*\hat{e}-q^*\hat{c})^{n}\Big]\hat{S}^{\dagger(N-n)}}{2N!}\\
        &=\sum_{k=0}^N\frac{\hat{S}^{N-k}Tr_E\Big[(p\hat{e}^\dagger+q\hat{c}^\dagger)^{k}|\Theta\rangle\langle \Theta|(p^*\hat{e}-q^*\hat{c})^{k}\Big]\hat{S}^{\dagger(N-k)}}{2N!}\\
        &=\sum_{k=0}^N\sum_{m=0}^k\frac{(C_k^{m})^2|p|^{2m}(-|q|^2)^{k-m}m!(k-m)!\hat{S}^{N-k}
        |0_s\rangle\langle 0_s|\hat{S}^{\dagger(N-k)}}{2N!}\\
        &=\frac{1}{2N!}\sum_{k=0}^Nk!(|p|^2-|q|^2)^k\hat{S}^{N-k}|0_s\rangle\langle 0_s|\hat{S}^{\dagger(N-k)}.
        \end{split}
        \end{equation}
 After the phase shift and reverse CPG operations, the output state of the hybrid system can be represented as the density matrix:
        \begin{equation}
            \hat{\rho}_L=e^{i\theta J_z\sigma_z}\rho_{1}e^{-i\theta J_z\sigma_z}.
          \end{equation}
 Furthermore, by tracing out the harmonic oscillator variables, the reduced density matrix of the spin state is finally obtained as:
        \begin{equation}
            \hat{\rho}_s=\rho_{00}|\downarrow\rangle\langle\downarrow|+\rho_{11}|\uparrow\rangle\langle\uparrow|
            +R|\downarrow\rangle\langle\uparrow|+R^*|\uparrow\rangle\langle\downarrow|,
        \end{equation}
 with
        \begin{equation}
          \begin{split}
          \rho_{00} &= Tr(\hat{\rho}_{00})\\
          &=\sum_{k=0}^N\sum_{l=0}^{N-k}\frac{(C_{N-k}^l)^2|\alpha_1|^{2l}|\beta_1|^{2N-2k-2l}l!(N\!-\!k\!-\!l)!k!(p^2+|q|^2)^k}{2N!}\\
          &=\frac{1}{2N!}\sum_{k=0}^N(|\alpha_1|^2+|\beta_1|^2)^{N-k}(p^2+|q|^2)^k(N-k)!k!=\frac{1}{2},
          \end{split}
        \end{equation}
 and
        \begin{equation}
          \begin{split}
          R &= Tr(\hat{\rho}_{01})\\
          &=\sum_{k=0}^N\sum_{j=0}^{N-k}\frac{(C_{N-k}^j)^2[C_N^k]^2\alpha_1^{2j}\beta_1^{2N-2k-2j}j!(N\!-\!k\!-\!j)!k!(p^2-|q|^2)^k}{2N!}\\
          &=\frac{1}{2N!}\sum_{k=0}[C_N^k]^2(N-k)!k!(p^2-|q|^2)^k(\alpha_1^2+\beta_1^2)^{N-k}\\
          &=\frac{1}{2}(\alpha_1^2+\beta_1^2+p^2-|q|^2)^N.
          \end{split}
        \end{equation}
 Above $\alpha_1=\alpha e^{i\theta/2}$,$\beta_1=\beta e^{-i\theta/2}$, and $\rho_{11}=1-\rho_{00}=1/2$, .
 Therefore,  the reduced density matrix of the spin state is :
          \begin{equation}
            \hat{\rho}_s=\frac{1}{2}(|\downarrow\rangle\langle\downarrow|+|\uparrow\rangle\langle\uparrow|)
            +\rho_{01}|\downarrow\rangle\langle\uparrow|+\rho_{10}|\uparrow\rangle\langle\downarrow|.
          \end{equation}
Thus Eq.~(\ref{rho_L}) can be obtained.
\end{appendix}

\end{document}